\documentclass[letterpaper,twocolumn,10pt]{article}
\usepackage{usenix,epsfig,endnotes,amsmath, authblk}
\begin{document}

\date{}


\title{\Large \bf Inference-to-complete: A High-performance and Programmable Data-plane Co-processor for Neural-network-driven Traffic Analysis }


\author[1]{Dong Wen}
\author[1]{Zhongpei Liu}
\author[2]{Tong Yang}
\author[1]{Tao Li}
\author[1]{Tianyun Li}
\author[1]{Chenglong Li}
\author[1]{Jie Li}
\author[1]{Zhigang Sun}

\affil[1]{National University of Defense Technology}
\affil[2]{Peking University}

\maketitle

\thispagestyle{empty}

\subsection*{Abstract}
Neural-networks-driven intelligent data-plane (NN-driven IDP) is becoming an emerging topic for excellent accuracy and high performance.
Meanwhile we argue that NN-driven IDP should satisfy three design goals: the flexibility to support various NNs models, the low-latency-high-throughput inference performance, and the data-plane-unawareness harming no performance and functionality.
Unfortunately, existing work either over-modify NNs for IDP, or insert inline pipelined accelerators into the data-plane, failing to meet the flexibility and unawareness goals.

In this paper, we propose Kaleidoscope, a flexible and high-performance co-processor located at the bypass of the data-plane.
To address the challenge of meeting three design goals, three key techniques are presented.
The programmable run-to-completion accelerators are developed for flexible inference.
To further improve performance, we design a scalable inference engine which completes low-latency and low-cost inference for the mouse flows, and perform complex NNs with high-accuracy for the elephant flows. 
Finally, raw-bytes-based NNs are introduced, which help to achieve unawareness.
We prototype Kaleidoscope on both FPGA and ASIC library.
In evaluation on six NNs models, Kaleidoscope reaches 256-352 ns inference latency and 100 Gbps throughput with negligible influence on the data-plane.
The on-board tested NNs perform state-of-the-art accuracy among other NN-driven IDP, exhibiting the the significant impact of flexibility on enhancing traffic analysis accuracy.

The source code is at https://github.com/Kaleidoscope-arch/Kaleidoscope.

\section{Introduction}

Intelligent data-plane (IDP) \cite{flowlet, IDP, nomos, IDPtc}, has become an emerging topic in the domain, which aims to equip data-plane with intelligent algorithms like decision trees \cite{dream, flowlens, zhaoshuang1} and neural networks (NN) for traffic analysis \cite{N3IC, Taurus}.
IDP is a networking-native platform for intelligent algorithms with low-latency-high-throughput.
Among various algorithms on IDP, NN is one of the most focused for high accuracy performance.
Therefore, NN-driven IDP receives attention from both the academic and industrial sectors \cite{broadcom, bos, N3IC}.

\subsection{Design Goals and Related Work}

To provide better traffic analysis service for networking, we argue that an NN-driven IDP should satisfy three design goals as following.
\textbf{(i) Flexibility to support different NN models.}
With ever fast-changing traffic and various deployment environments, only supporting NNs with specific types and sizes cannot efficiently enable traffic analysis.
Meanwhile, flexibility of NNs support provides better choice for different cases.
For example, light-weight multi-layer perceptron (MLP) \cite{mlp, mlp1} models are suitable for fast analysis, while convolutional neural networks (CNN) \cite{fenxi, zeroday} can perform accurate inference for accuracy-sensitive tasks.
\textbf{(ii) Low-latency and high-throughput inference performance.} 
Data-plane has achieved the performance with over 100 Gbps throughput and hundreds of nano-seconds (ns) forwarding latency \cite{trio}.
The inference of NN-driven IDP should match this performance requirement.
\textbf{(iii) Ensuring no significant harm on the performance and function of data-plane, namely data-plane-unawareness.}
The primary task of the data-plane is to operate correct and high-performance data forwarding. 
Hence, NN inference on IDP must not impact the functionality and performance of the data-plane \cite{wugantc}.
We noted this as data-plane-unawareness (DP-unawareness) in this paper.

Unfortunately, existing work on NN-driven IDP fail to satisfy these goals at the same time.
Related work can be categorized by the software and the hardware solutions.
The former aim to modify NN models for current IDP platforms \cite{nomos, nomos22, bos}.
For instance, BoS \cite{bos} compresses the parameter of NNs models into one-bit format \cite{BNN}, and replaces complex inference computation with look-up table on programmable switch.
Although BoS achieves line-rate throughput, such software solution endures accuracy loss and only investigates on specific types of NNs, lacking the flexibility to support other model types.
Also, its massive TCAM consumption may influence other network functions co-located on the device \cite{menshen}.
While in the existing hardware solutions \cite{N3IC, Taurus}, pipelined accelerators are developed and integrated within the forwarding pipeline via an inline way, which maintains line-rate throughput and leverages data-plane to extract feature as input.
However, such pipelined accelerators sacrifice flexibility because accelerators cannot support NN models that exceeding pipeline stages, nor it will block the forwarding pipeline by loop \cite{p4, rmt, drmt}. 
Moreover, these inline hardware integration are constrained to small pipelined accelerators, as larger ones would significantly increase latency, thereby further compromising model flexibility.
In fact, about 45\% of latency is introduced by NN inference in Taurus \cite{Taurus}.
In conclusion, related work on NN-driven IDP cannot meet the flexibility and DP-unawareness requirements.

\subsection{Challenges and Our Solutions}

To address the problem, our general idea is to propose an NN co-processor located at the \textbf{bypass}\footnote{Bypass here has slight difference with "bypass" in kernel bypass which removes operations from the kernel space, while "bypass" in this paper means our co-processor is parallel with the main forwarding pipeline.} 
of the data-plane, which is essentially different from the related work.
A bypass NN co-processor can help to meet all design goals by (i) removing the limitation from the forwarding pipeline for flexible inference architecture, (ii) setting asynchronous clock domains from the data-plane for optimized running frequency and performance, (iii) isolating inference and forwarding for DP-unawareness.

However, it still faces significant challenges to develop a bypass co-processor. 
\textbf{Firstly, how to acquire both flexibility and high-performance?}
Pipeline architectures are high-performance but rigid, while run-to-completion (RTC) architectures have better flexibility but fall short on performance.
It is difficult to design architecture acquiring both goals.
\textbf{Secondly, how to remove NNs dependency on the data-plane for DP-unawareness.}
Current NN models on IDP heavily rely on the data-plane for feature extraction, which increases latency and consumes massive resource, violating the DP-unawareness.
However, designing a specific feature extractor on the bypass involves heavy engineering workload.
These difficulties impede the design and implementation of the bypass co-processor.

In this paper, we propose \textbf{Kaleidoscope}\footnote{Kaleidoscope is an optical mirror reflecting various image and colour.}, a flexible, programmable and high-performance co-processor for NN-driven IDP on the bypass, which can meet three design goals simultaneously.
Three key techniques are presented in Kaleidoscope.
\textbf{Key 1: Programmable RTC architectures for flexible inference.}
Computation across various NNs can be decomposed into the combination of generic matrix-vector multiplication (GEMV), generic matrix-matrix multiplication (GEMM) and non-linear activation functions. 
We design our programmable RTC accelerators towards basic operations above, enabling GEMM/GEMV with different dimensions through blocked multiplication and RTC looping, thereby supporting NN models flexibly.
\textbf{Key 2: High-performance inference engine towards elephant/mouse flows.}
As compared in the table \ref{elephant}, elephant and mouse flows demand differential requirements on inference.
Specifically, mouse flows need small NNs for fast inference while elephant flows require larger NNs for accurate inference.
Leveraging this observation, two kinds of RTC accelerator are developed in our inference engine, namely the fast process element (FPE), which completes low-latency and low-cost inference with smaller NNs, and the heavy process element (HPE), which can support complex NNs with high throughput.
Moreover, both accelerators can be scaled with larger parallelism, further improving computational performance.
\textbf{Key 3: Raw-bytes-based NNs for DP-unawareness.}
To remove the data-plane-dependency for traffic feature extraction in other IDP work, we introduce raw-bytes-based NNs \cite{etbert, 1dcnn, beiyouiwqos} in Kaleidoscope, which directly feeds raw packets into NNs for inference.
With traffic being mirrored to the bypass, Kaleidoscope utilizes no function from the data-plane, achieving full DP-unawareness.

Kaleidoscope is implemented by over 7000 lines of Verilog HDL codes on our FPGA-based test-bed.
Our co-processor can efficiently support six NN models including MLP, CNN and recurrent NNs (RNN) with 256-352 ns inference latency and 100 Gbps throughput on-board, exhibiting both flexibility and high-performance.
Under the 28nm ASIC library, Kaleidoscope further performs 1.6 Tbps throughput.
Simultaneously, the entire data-plane only experiences extra five-clock-cycles latency due to querying inference result, achieving the goal of DP-unawareness.
It is also noteworthy that on-board tested models acquire the state-of-the-art (SOTA) accuracy of 92.8\%-99.2\% compared with other NN-driven IDP, reflecting the accuracy improvement from flexible NNs support.

\subsection {Contributions}
The contribution of this paper is three-fold as below:

$\cdot$ We propose Kaleidoscope, the first bypass co-processor for NN-driven IDP, which acquires design goals of flexibility, high-performance and DP-unawareness.
We open-source our code, including over 7000 lines Verilog HDL codes and 1000 lines Python codes for hardware, tool-chains and NNs models.

$\cdot$ We present a systematic hardware design, including flexible RTC architectures for inference, a high-performance engine towards elephant/mouse flows and introducing raw-bytes-based NNs with traffic mirror.
It successfully combines hardware, networking and NNs models into the proposed architecture and meets all design goals.

$\cdot$ We established a FPGA-based 100 Gbps test-bed to validate Kaleidoscope.
The evaluation proves that Kaleidoscope satisfies all design goals, with on-boarded tested NNs achieve SOTA performance compared with other NN-driven IDP.

\section{Background, Preliminary and Motivation}

\subsection {NN Models for Traffic Analysis}

\textbf{Flexible NN Models.}
To better cope with real-world network traffic, NN models are keeping evolving on architectures and model sizes.
For example, transformer \cite{etbert, mtc, mtt, newhope} and Mamba \cite{mamba} models are latest NN models introduced in traffic analysis.
It is necessary for NN-driven IDP to support flexible and fast-changing NN models, however, existing solutions cannot satisfy the flexibility requirement.

\textbf{NN Models on IDP.}
Currently, most of the NNs on IDP take statistic traffic features \cite{tongji1, tongji2} as model input.
For instance, the RNN in BoS relies on packets length and packets arrival interval for flow-level inference.
However, the extraction of traffic features occupies function and resources on the data-plane, which influences other co-located network functions.
Moreover in the hardware solutions, to borrow the feature extraction from the data-plane, the pipelined accelerators have to be involved with inline integration with the forwarding pipeline.
This brings in-optimal latency and limited hardware design space.

\textbf{Raw-bytes-based NN Models.}
Instead of adopting statistic traffic features, raw-bytes-based NNs \cite{etbert, 1dcnn, beiyouiwqos} directly feed models with raw packet for inference.
This type of NNs can utilize the information of protocol (especially the application-level protocol) and data distribution carried in the packet for accurate inference, removing the dependency on the data-plane for feature extraction.
Meanwhile, the size of raw-bytes-based NN models is usually larger than feature-based ones, which is challenging to deploy on previous IDP.
However, the flexibility design of our RTC architectures makes it feasible to implement raw-bytes-based NNs on the data-plane.

\subsection {Elephant Flows and Mouse Flows}
Elephant flows and mouse flows are phenomena widely discussed where small numbers of elephant flows occupy the majority of the bandwidth, while a larger number of mouse flows occupy smaller data volume.
Intrinsically, such phenomena also fit in the cases of traffic analysis.
As listed in the table \ref{elephant}, we notice that elephant flows and mouse flows demand differential requirements on NN inference.
Mouse flows have fewer packets and shorter duration time, requiring small NNs with fast inference.
Meanwhile, elephant flows carry more packets, occupying larger bandwidth volume and longer lasting time.
Therefore, complex NN models with higher accuracy are necessary for analysing elephant flows.

\textbf{By distinguishing mouse flows and elephant flows, we can conduct traffic analysis with balanced inference costs, achieving high-performance design goal.}
In Kaleidoscope, two kinds of RTC accelerators: FPE and HPE, are designed towards elephant flows and mouse flows respectively.
Smaller but faster NNs models are deployed on the FPE, efficiently processing the most proportion of network flows with lower cost.
On the other hand, HPE can complete inference of more complex NNs to acquire high-accuracy for elephant flows.
Both accelerators can flexibly support various NN models, meeting both flexibility and high-performance goals.
\begin{table}[htbp]
\begin{center}
\begin{tabular}{|c|c|c|}
\hline
\textbf{Characteristic} & \textbf{Elephant flows} & \textbf{Mouse flows} \\
\hline
Volume & Large & Small \\
\hline
Duration time & Long & Short \\
\hline
Amount & Small & Large \\
\hline
Inference req. & High-accuracy & Low-latency \\
\hline
\end{tabular}
\caption{Elephant flows and mouse flows}
\label{elephant}
\end{center}
\end{table}

\subsection {GEMV/GEMM on RTC Accelerators}
\label{sec:2.3}
\begin{figure}[htbp]
\centerline{\includegraphics[width=1\linewidth]{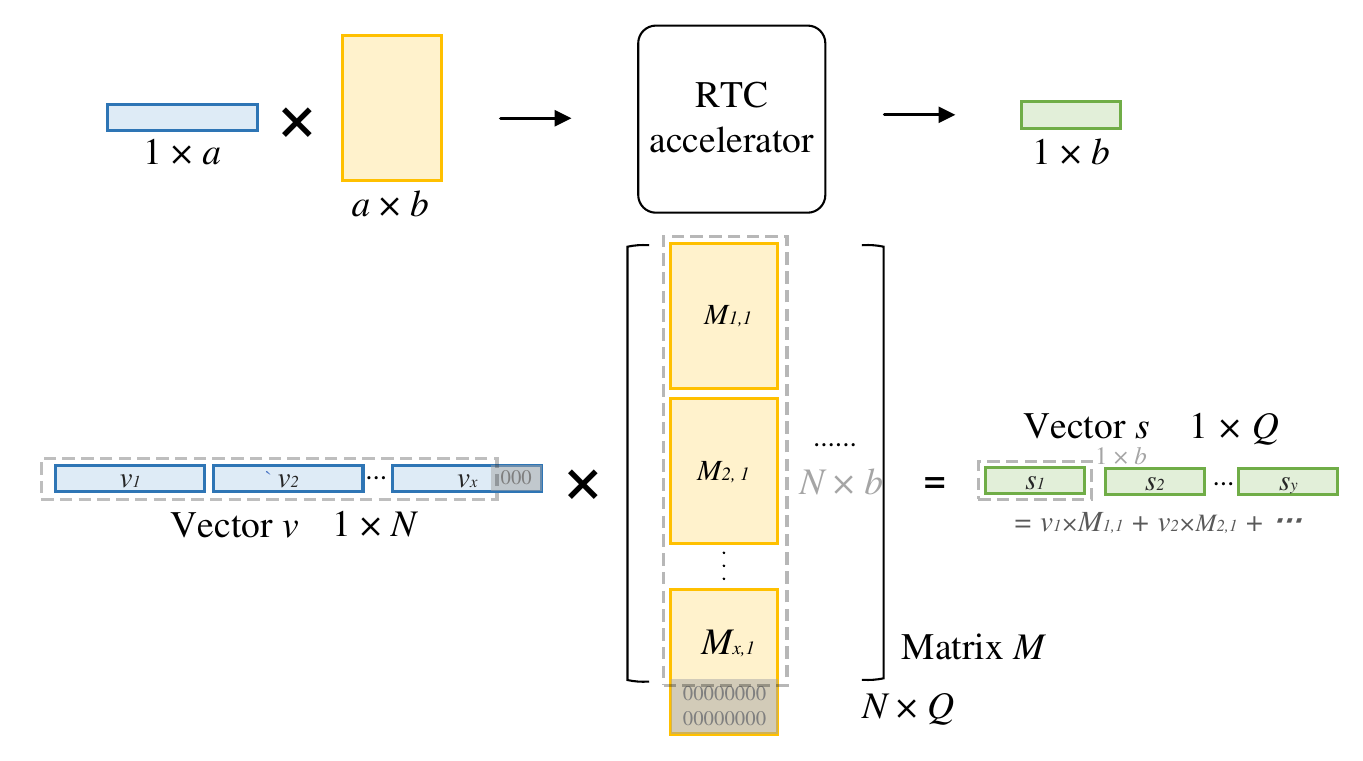}}
\caption{Blocked GEMV on the RTC accelerator.}
\label{GEMV}
\end{figure}

Different sizes and types of NNs models can be decomposed into GEMV/GEMM and non-linear activation, where GEMV/GEMM operations occupy the most of the computation.
The main idea of Kaleidoscope is to deign RTC accelerators for variable dimensions GEMM/GEMV, thereby supporting flexible NNs models.
In the figure \ref{GEMV}, we take GEMV as an example to illustrate how RTC architectures achieve flexibility.

Considering we already have a RTC accelerators which can complete GEMV with the dimension of $(1, a) \times (a, b)$ in a macro-cycle.
At the same time, one layer in a NN model is mapped into the GEMV of $(1, N) \times (N, Q)$ between the vector $v$ and the matrix $M$.
Apparently, the RTC accelerator cannot directly support such GEMV operations.
However, we notice that $v$ can be divided into $x$ segments of sub-vector, whose dimension is $(1, a)$.
Such sub-vectors are noted as $v_i$, and the tail part of the $v$ is padded with zero for alignment.
Similarly, the matrix $M$ is divided into $x \times y$ blocks of sub-matrix $M_{i, j}$, whose dimension is $(a, b)$.
The zero-padding operation is conducted on the corner blocks as well.
Following the below equation \ref{block}, the result vector $s$ is accumulated by the temporal results from multiple multiplications $v_i \times M_{i, j}$, which can be mapped to the RTC accelerator seamlessly.
The accumulation operation can be completed by accumulators supplemented in the RTC module.
Following above computing method, the RTC architecture is able to perform GEMV with variable dimensions, thereby supporting NNs models flexibly.
In the cases of GEMM, extending the vector $v$ into a matrix with $(P, N)$ dimension, it acquires the similar computing method for flexible inference.

$$
\begin{aligned}
s & =v \cdot M \\
& =\left(v_{1}, v_{2}, \cdots v_{x}\right) \cdot\left(\begin{array}{ccc}
M_{1,1} & \cdots  &M_{1,y} \\
\vdots & & \vdots \\
M_{x,1} & \cdots & M_{x,y} \\

\end{array}\right) \\
& =\left(\sum_{i=1}^{x} V_{i} M_{i,1}, \cdots, \sum_{i=1}^{x} M_{i,y}\right)
\end{aligned}
\label{block}
$$

Based on the illustration above, it can be inferred that blocked GEMV/GEMM demands loop computing and temporal data storage.
\textbf{However, when integrated inside the data-plane, accelerators proposed by previous work cannot support the loop operation, otherwise it will block the forwarding pipeline.
This is the essential reason why the existing hardware solutions have significant faults on flexibility.}
On contrast, the proposed RTC architectures can flexibly support loops for various NN inference.

\section{Architecture}


\subsection{Overview and Workflow}

\begin{figure*}[htbp]
\centerline{\includegraphics[width=1\linewidth]{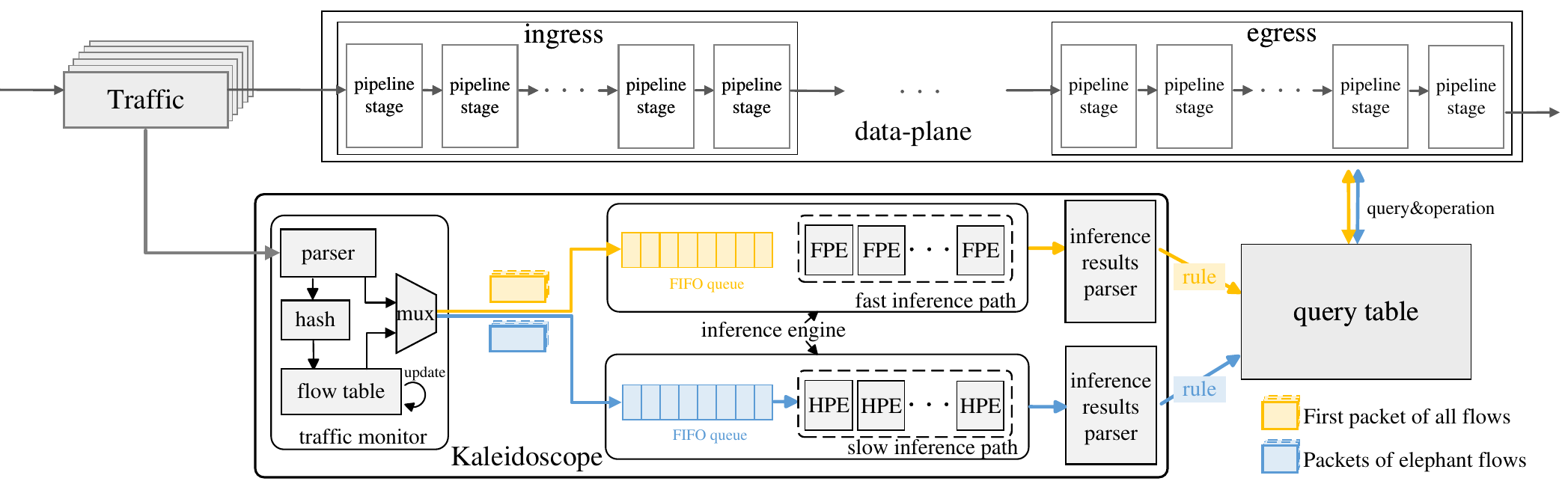}}
\caption{The architecture of Kaleidoscope co-processor.}
\label{arch}
\end{figure*}
 
The architecture of entire co-processor is depicted in the figure \ref{arch}.
The start of the co-processor is the traffic monitor, which distinguishes elephant flows and mouse flows, determines the first-appearing flows and calculates hash value for each flow.
Kaleidoscope utilizes this hash value to identify different flows.
Packets in elephant flows and mouse flows are sent to separated inference path, which are called the fast inference path and the slow inference path.
Multiple FPEs and HPEs, which adopt programmable and scalable RTC architecture, are wrapped in these path respectively.
FPE focuses on accelerating GEMV operations, completing small NNs inference within hundreds of nano-seconds.
HPE performs GEMM with optimal throughput, being able to perform inference of complex NN models.
Auxiliary components like \texttt{argmax} in parser module are appended behind FPEs and HPEs.

It is challenging to design the workflow for Kaleidoscope because it takes time to collect the flow size, where traffic may escape before distinguishing the elephant and the mouse flows.
To address the problem, we put the first packet of all flows to the fast inference path without determining the flow size.
When arriving the volume threshold, the elephant flows will be analysis by the complex NN models in the slow inference path for second time.
Inference results are designed to interact with the data-plane only via a query table.
The proposed workflow provides full coverage of traffic and DP-unawareness for NN-driven IDP.


\subsection{Traffic Monitor}
The traffic monitor is developed for three functions: distinguishing the elephant and mouse flows, identifying the first-appearing flows, and calculating hash value for each flow.
Several dimensions are utilized to distinguish the elephant and mouse flows, during time, packets counts and traffic volume for instance \cite{elephant1, elephant2}.
For convenience, this paper chooses packets counts in a flow to separate two types of traffic.

The architecture of traffic monitor is depicted in the figure \ref{arch}, which consists a parser to extract IP tuple and valid bytes for raw-bytes-based NNs.
It is also designed with a hash calculator (Hash) and a flow table.
The information of first-appearing flag and packet counts are saved in the flow table, which is indexed by the calculated hash value.
The hash value is calculated from IP tuple for identifying a flow.
Therefore, when new traffic arrives, the traffic monitor will check whether it is a first-appearing flow by querying the flow table, then updates the packet counts.
If the packet counts reaches the threshold of the elephant flows, corresponding bytes are sent to the slow inference path via MUX module.

\subsection{Fast Process Element}

Fast processing element is a programmable RTC accelerator designed for low-latency and flexible inference on the mouse flows.
In the cases of small NNs with low-latency inference requirement, the most suitable batch-size for inference is 1, which usually corresponds to the GEMV operations.
Therefore, the kernel of FPE is to accelerate the blocked GEMV computation.
However, the accumulation and data loading can be the latency bottleneck in blocked GEMV.
To solve the problem, we present several key designs for low-latency performance of FPE.

\subsubsection{Low-latency Architecture for GEMV}

\begin{figure}[htbp]
\centerline{\includegraphics[width=1\linewidth]{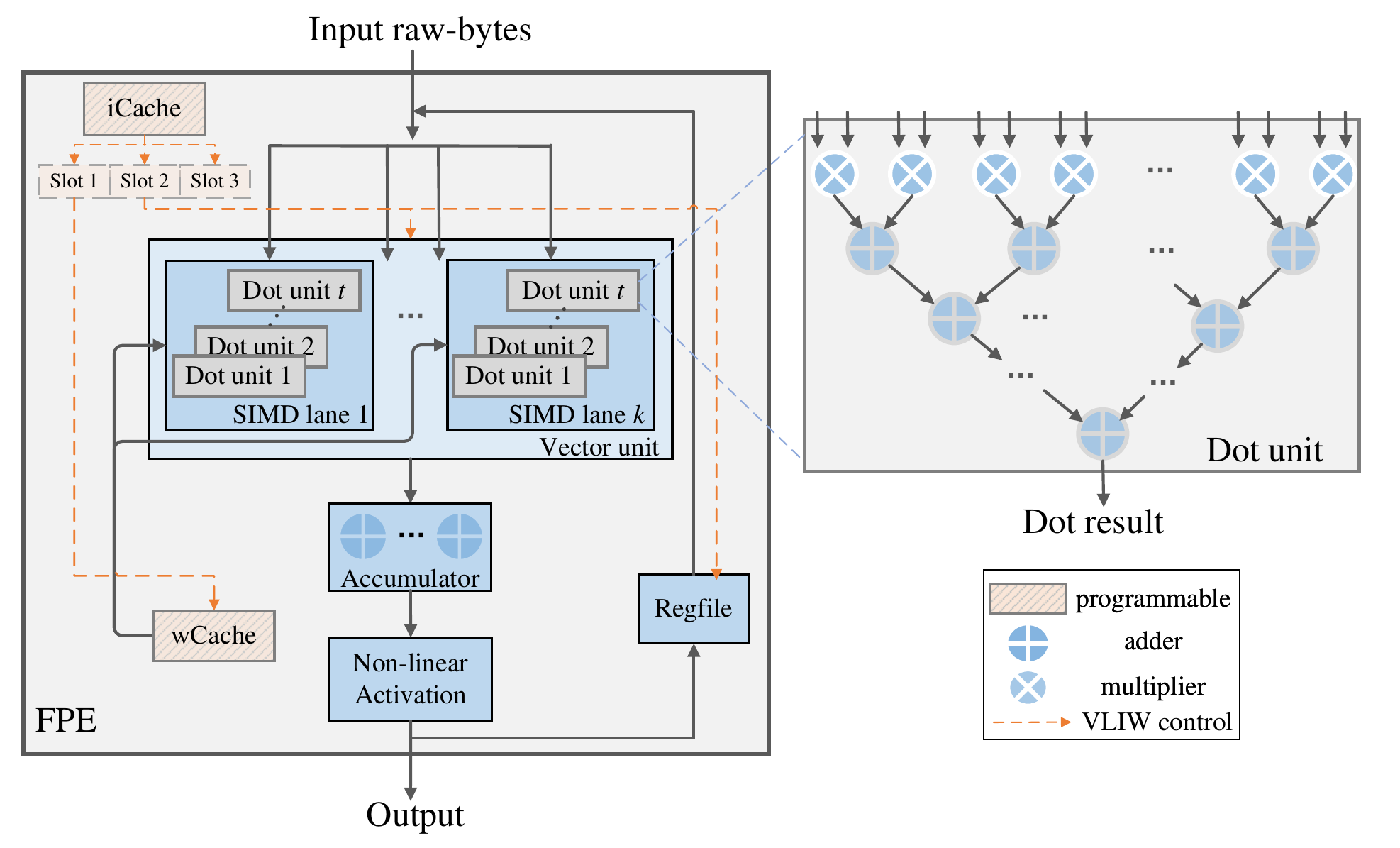}}
\caption{The architecture design of FPE.}
\label{FPE}
\end{figure}

\textbf{Optimal dot unit.}
The architecture of dot unit is shown in the figure \ref{FPE}.
There are $n$ parallel multipliers inside the dot unit, followed by an adder tree to accumulate multiplication results to the dot value.
The proposed dot units can complete the dot operation between two $(1, n)$ vectors with the optimal latency complexity of $O(logn)$.

\textbf{Hierarchy and inline accumulator design.}
Stacking $t$ dot units via a single-instruction-multiple-data (SIMD) way, it gets the SIMD lane in the figure \ref{FPE}, which can operate the GEMV operation between a $(1, n)$ vector and a $(n, t)$ matrix for final $(1, t)$ output.
The computation in SIMD lane hierarchy needs no accumulation because dot units inside a SIMD lane share the same input vector and operate on dependant columns of the matrix.
To further extend the computational dimensions, multiple SIMD lanes are wrapped inside the FPE.
Thus FPE can compute GEMV between a $(1, n\times k)$ vector and a $(n \times k, t)$ matrix, $k$ stands for the amount of the SIMD lanes.
However, SIMD lanes receive different segments of the ($n \times k$)-length input vector, where accumulators are needed for merging the temporal results from blocked GEMV.
To optimize the merging latency in GEMV, we design inline accumulators following the hierarchy of SIMD lane.
Temporal data calculated by SIMD lanes can be directly sent to the accumulators, reducing the accumulation latency.

\textbf{Parallel data access.}
To reduce the data-loading latency, we decouple the data-path for parallel data-loading.
We separate the data-loading process into three dependent path, which are temporal-data path, instruction-fetching path and NN-parameter path, which correspond to the register file (Regfile), instruction cache (iCache) and parameter cache (pCache).
As discussed before, small NNs are employed for the mouse flows, where the size of temporal data and parameters fit in the on-chip storage.
Leveraging this characteristic, we develop regfile for saving the temporal data, providing low-latency data access for GEMV. 
iCache and pCache stores the instructions and parameters utilized by the inference.
Three types of data-loading are separated and parallel, which will not impact each others or block the inference.

\textbf{Very-long-instruction-word architecture for RTC.}
The FPE architecture design exhibits parallelizability, where GEMV computation, temporal data access and NNs parameter loading work without locking.
Utilizing this feature, we employ the very-long-instruction-word (VLIW) architecture for FPE to implement hardware parallelism for lower latency.
The proposed VLIW architecture has three instruction slots, namely the computation slot, parameter loading slot and the temporal data access slot, which control the blocked GEMV computation, the address for parameter loading and the read/write operations on regfile.
In each cycle, three instructions are issued by slots to control the data-flow of the FPE.
With proper instruction assignment, FPE can perform GEMV with the non-stall data flow, achieving low-latency inference.

\subsubsection{Flexibility for NNs Inference}
The FPE design above provides the low-latency capacity for GEMV computations with vaiable dimensions.
However, it demands further development to fully support flexible NNs inference.

\textbf{Quantized activation module.}
Activation is another important operation in NNs inference, which usually employs non-linear functions like \texttt{ReLU} and \texttt{Sigmod}.
We locate activation module following the accumulators, directly receiving GEMV results and saving data reloading.
Considering the complexity of non-linear functions, look-up table is leveraged to implement the activation operations with NNs quantization to reduce latency and resource consumption.
For example, given an 8-bit quantization data format, look-up table only occupies a 256-depth table.

\textbf{Instruction Set for FPE and programmable inference.}
The flexible NNs support demands RTC architectures can be programmed by the users.
Therefore, we design an instruction set for FPE programming, consists instructions for GEMV, activation and data accessing.
Controlling instructions like \texttt{FIN} and \texttt{START} are also presented to control the workflow.
Entire instruction set of FPE is listed in the appendix.
Users can upload their program into the iCache, and upload NNs parameters into pCache.
Both of the two cache can be updated run-time, providing full programmability.

\subsubsection{Scalability of FPE}
The FPE architecture can be scaled on two dimensions: 
(i) Extending the parallelism inside the FPE, like adding the amount of the SIMD lanes or increasing the size of the dot units. 
However, such method introduces under-utilization, because NNs models employed on the mouse flows are relative small in Kaleidoscope. 
(ii) Extending the parallelism by adding multiple FPEs, which is the method adopted by the Kaleidoscope.
We assign inference tasks of different flows on each FPE, therefore multiple FPEs can work dependently without locking, exhibiting scalability.

\subsection{Heavy Process Element}
\begin{figure}[htbp]
\centerline{\includegraphics[width=1\linewidth]{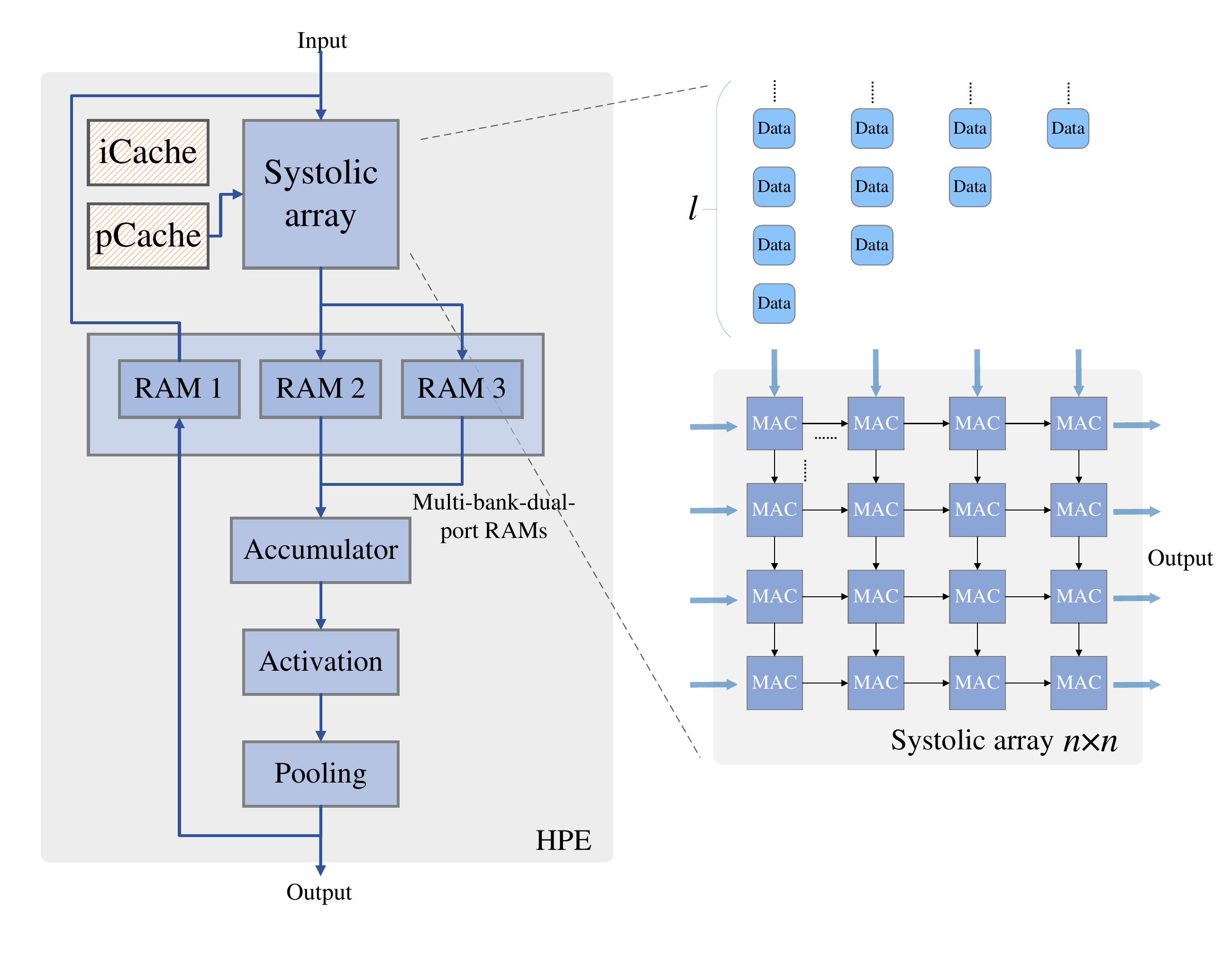}}
\caption{The architecture of HPE.}
\label{HPE}
\end{figure}

Heavy process element is a programmable RTC accelerators designed for inference of complex NNs models on the elephant flows.
The inference of complex NNs are usually mapped to the GEMM for higher computational complexity.
Therefore, the kernel of HPE is based on systolic array \cite{array}, a throughput-optimal architecture for GEMM computation adopted by Google TPU \cite{tpu}.

However, the size of temporal data in the blocked GEMM is larger than GEMV, which demands more memory component and is difficult to be completed by the inline accumulators.
This further introduces the dilemma of data reloading and ports conflict on the memory, which may block the data flow of the GEMM computation.
To address the problem, we propose several designs for high-efficient inference.

\subsubsection{The Introduction of Systolic Array}
As shown in the picture \ref{HPE}, systolic array employs multiply-adder (MAC) for computing, which can conduct $d = a\times b + c$ by one cycle.
The $c$ transferred between MAC is utilized for sum in dot operations.
A $n \times n$ sizes systolic array performs GEMM between a $(l, n)$ matrix and a $(n, n)$ matrix with $O(n+l)$ latency complexity.
Systolic array is able to support GEMM with variable dimensions following the blocked GEMM methods in section \ref{sec:2.3}.

\textbf{Analysis on data flow.}
It needs three data access ports for systolic array to function without stall: two ports for reading and one port for writing back.
In the cases of NNs inference, one of the data reading carries NNs parameters, thereby can be stored in pCache by read-only mode. 
The other reading port and the writing port operate on temporal data of the inference.
From the view of the accumulator, however, it also demands at least two ports for reading and writing respectively.
In fact, considering accumulation takes two data for input and output one merged data, two reading ports and one writing ports can justly match the computing speed of the systolic array, otherwise it becomes the bottleneck of the GEMM operations and introduce stalls into the accelerator.

\subsubsection{Non-stall Data Flow Organization}
The original data flow of blocked GEMM on the systolic array can be divided into five types:
(1) Systolic array fetching read-only parameters.
(2) Systolic array fetching source data for multiplications between sub-matrices.
(3) Systolic array writing back the multiplication results.
(4) The accumulator fetching the two groups of multiplication results for merging.
(5) The accumulator writing back the merged data.
The data flow above involves with four reading operations and two writing operations which exhibits apparent dependency.

To implement the non-stall GEMM architecture, we propose to cut the intertwined data flows by the multi-bank-multi-port memory design.
Unlike small NNs supported by FPE, complex NNs inside the HPE demand larger on-chip memory space for computing and parameter saving.
Instead of simply increasing the space of memory, we divide the needed volume into three RAM blocks with dual read/write ports on each, as shown in the figure \ref{HPE}.
Therefore, the accelerator is equiped with six dependent data ports for both reading and writing.
Additionally, the NNs parameter is stored in pCache while the instructions are stored in iCache, which is the same with FPE.

The data flow is cut by following procedures.
First, systolic array receives parameters from the pCache and source data from the RAM 1, corresponding to the data flow 1 and flow 2.
The multiplication starts, with the temporal data being written back to the RAM 2 and RAM 3 by turns (flow 3).
At the same time, the accumulator fetches temporal data in parallel from RAM 2 and RAM 3, completes the accumulation for blocked GEMM (flow 4), and writes final results into RAM 1 (flow 5).
We separate five data flows by three dual-ports RAM blocks.
Due to the function characteristic of dual-ports RAM, systolic array and the accumulator can access all RAM blocks simultaneously, enabling the non-stall data flow for blocked GEMM.
Similarly, VLIW architecture is applied in HPE, which controls the data access and the computation hardware parallelism.

\textbf{Activation and pooling.}
The activation module in the HPE receives the data from accumulators, which is same with the FPE.
However, there exists other non-linear operations in complex NNs models, such as \texttt{maxpool} in CNNs.
For fully supporting various NNs models, we implement pooling modules behind the accumulator  and the activation data flow.


\subsection{Programming Interface and Instruction Set}
\subsubsection{Programming Interface}
\textbf{Hardware programming interface}
iCache and pCache in the RTC accelerator provide programmability support for Kaleidoscope.
Although two cache are set as read-only mode during inference, their data can be programmed by users in run-time.
Two cache can be modified via specific configuration traffic, or be programmed via other hardware methods like flashing memory.

\begin{figure}[htbp]
\centerline{\includegraphics[width=1\linewidth]{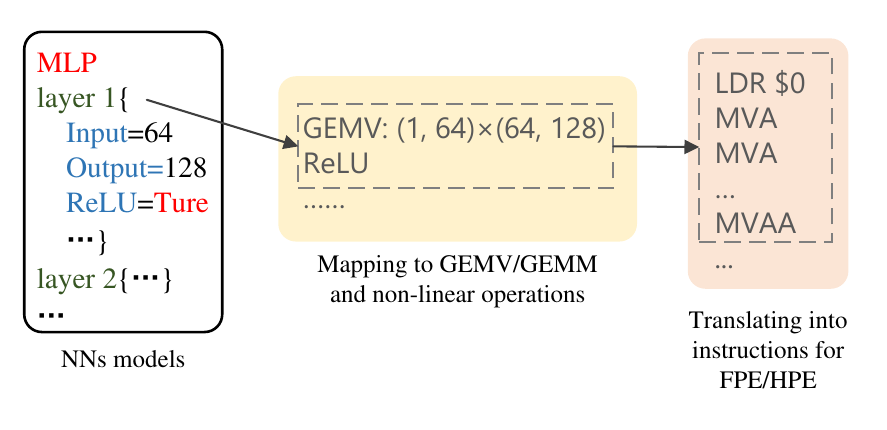}}
\caption{The workflow of programming NNs models.}
\label{userface}
\end{figure}

\textbf{Software programming interface.}
The software programming interface is established on decomposing inference into GEMV/GEMM and non-linear operations.
We take a MLP model in the figure \ref{userface} as example.
There are multiple NN layers stacked in the MLP with different sizes of input and output.
Layer 1 receives 64 dimensions of raw-bytes and generates 128 dimensions of output with \texttt{ReLU} activation, which can be mapped into a GEMV between a $(1, 64)$ vector and a $(64, 128)$ matrix, \texttt{ReLU} activation follows the GEMV operation.
This combination are further mapped on the RTC FPE hardware, coded by the instruction set of FPE.
We develop the scripts to automatically map NNs into GEMV/GEMM operations, and translate operations into instructions.

\subsubsection{Instruction Set}
To program the RTC accelerators for flexibly supporting NNs models, we design two instruction sets for FPE and HPE respectively.
The most important instruction for FPE is \texttt{MV}, which stands for the GEMV operations. 
Also, MV-Accumulate (\texttt{MVA}) and MV-Accumulate-Activation (\texttt{MVAA}) are designed for utilizing inline accumulator and activation module.
Similarly in HPE, the instruction set is designed towards GEMM, which consists Matrix-Multiplication (MM) for sub-matrix, Accumulate (\texttt{ACC}) for the accumulator and Accumulate-Activation (\texttt{ACCA}), Accumulate-Pool(\texttt{ACCP}) for using inline activation and pooling modules. 
The data access instructions like Load-Regfile (\texttt{LDR}) and Load-Parameter (\texttt{LDP}) are also designed in our instruction sets.
The entire instruction sets are listed in the appendix A.4.

\section{Implementation and Test-bed}

\begin{figure}[htbp]
\centerline{\includegraphics[width=0.87\linewidth]{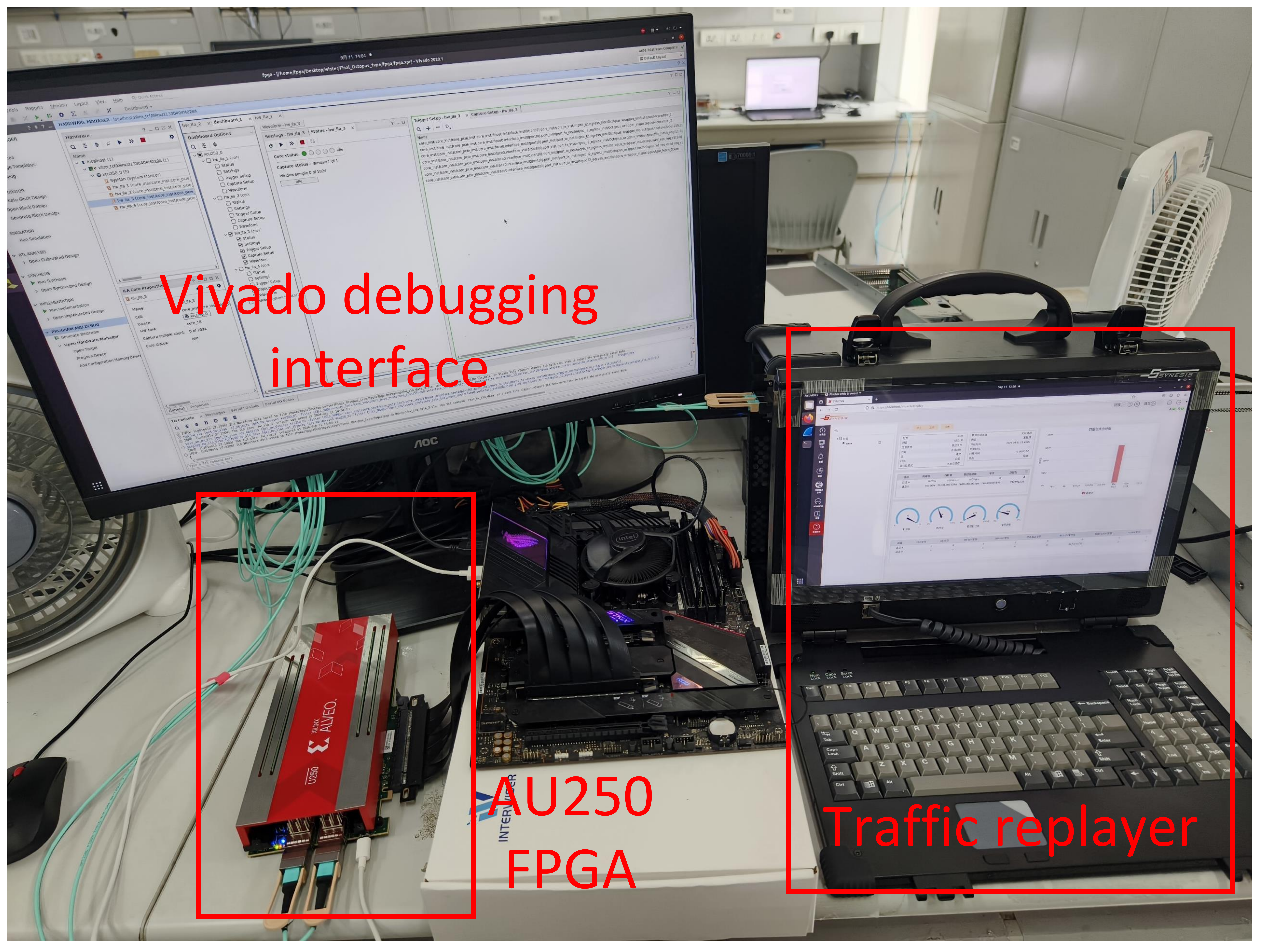}}
\caption{100Gbps test-bed for Kaleidoscope.}
\label{testbed}
\end{figure}



\subsection{Hardware Implementation}
\begin{figure}[htbp]
\centerline{\includegraphics[width=1\linewidth]{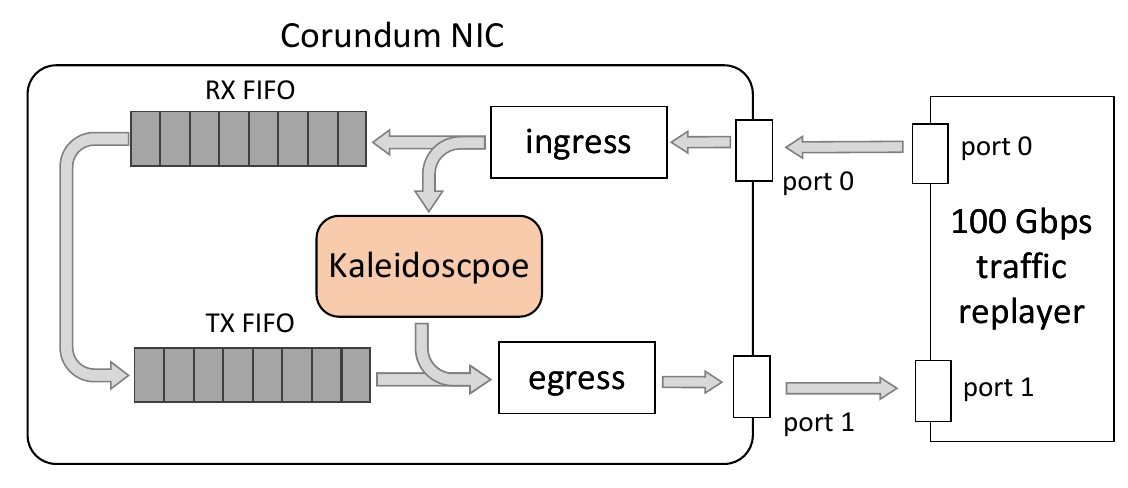}}
\caption{Integrating Kaleidoscope on the bypass of the 100 Gbps NIC.}
\label{NIC}
\end{figure}

We develop Kaleidoscope co-processor by over 7000 lines of Verilog HDL codes.
The Kaleidoscope is implement based on Corundum \cite{corundum}, an open-source 100 Gbps NIC platform.
The aim of FPGA-based implementation is to evaluate the capacity of Kaleidoscope of handling high band-width, therefore, we do not use the NIC-to-host function path and modify the data flow inside the Corundum with traffic loop test by the figure  \ref{NIC}.
The NIC-to-host path in the original Corundum is still kept to maintain the completed functionality of the NIC systems.
The input traffic from \texttt{port 0} are directly sent to the egress of the \texttt{port 1}, with Kaleidoscope receiving the mirrored traffic and writing inference results into the query table at the bypass.
The top of the Kaleidoscope codes and the query table are inserted at \texttt{m2\_egress.v} file of the Corundum project, where we develop a bypass interface for NNs inference and generating a new asynchronous clock for Kaleidoscope.
The hardware platform for implementation is Xilinx AU-250 FPGA, the implement runs on the Vivado 2020.1 with the \texttt{NetDelayLow} implement strategy option.

\textbf{Kaleidoscope configuration.}
We set following configurations for Kaleidoscope.
In traffic monitor, we choose 16 as the packet count threshold for distinguishing the elephant and the mouse flows.
Besides, we reuse the \texttt{rx\_hash.v} in the Corundum as the hash calculator, and set the depth of the flow table and query table as 64K.
We also sets a 512-depth FIFO queue for each FPE.
A FPE has 4 SIMD lanes, each contains 8 dot units.
One dot unit is equipped with 8 parallel multiplier and 7 adders.
Thereby one FPE can perform GEMV between a $(1, 32)$ vector and a $(32, 8)$ matrix in a macro-cycle.
A Regfile with 256-bit$\times$32 sizes together with a 8KB pCache and a 1KB iCache is developed in FPE.
HPE is implemented based on a $32 \times 32$ systolic array, with a 8 KB iCache, a 512KB pCache and three 256-bit$\times$1024 dual ports RAM blocks.
All computing components are designed for quantized Fix-8 data format, which is further discussed in section \ref{sec:5.2.3}.

To evaluate the scalability of Kaleidoscope, we implement three main versions, namely K-Base, K-4FPE and K-8FPE, which consists one HPE plus one FPE, two FPE and four FPE respectively.
Kaleidoscope is equipped with more FPE than HPE because the amount of the mouse flows is larger than the elephant flows.
Additionally, we also develop a inline version of K-Base (Inline in the tale \ref{FPGA}) for comparison.
All implementation results on FPGA are in the table \ref{FPGA}.

\textbf{Analysis on inline and bypass implementation.} The bypass architecture of Kaleidoscope can efficiently isolate the clock domain of the data-plane and co-processor, maintaining the 322Mhz running frequency and 100Gbps line-rate of the forwarding pipeline.
The frequency of Kaleidoscope (250/223/200Mhz) is lower than Corundum NIC because the parallelism of inference hardware is far more larger than the data-plane, which increases the difficulty of routing and implementation on the FPGA.
On contrast, the inline accelerator integrated inside the data-plane shares the same clock domain with the Corundum NIC.
Such architecture makes it more challenging to optimize on FPGA and finally lowering the running frequency of the entire system by 37.9\%, which corresponds to 59.6 Gbps forwarding performance on the Corundum.
Apparently, the inline architecture directly harms the performance of the data-plane, failing to meet the DP-unawareness requirement.

As to the resource utilization, Kaleidoscope consumes substantial DSP for computation, however, it will not conflict with the Corundum data-plane because only one DSP is utilized in the Corundum.
Similarly, Kaleidoscope introduces acceptable consumption on LUT logic and Block RAM (BRAM), which will not impact the performance and the functionality of the data-plane.
The implementation results on hardware frequency and resource consumption exhibit the advantages of bypass Kaleidoscope architecture on DP-unawareness.

\begin{table}[htbp]
\begin{center}
\begin{tabular}{|c|c|c|c|c|}
\hline
Architecture & LUT & BRAM & DSP & Freq. (MHz)  \\
\hline
\textbf{K-Base} & 7\% & 20\% & 16\% &286/322\\
\hline
\textbf{K-4FPE} & 10\%& 27\%& 25\%& 250/322\\
\hline
\textbf{K-8FPE} & 13\%& 33\%& 38\%& 250/322\\
\hline
NIC-base & 5\% & 14\% & 1\% & 322\\
\hline
Inline & 13\%& 20\% & 16\% &200\\
\hline
\end{tabular}
\caption{Kaleidoscope implementation results on FPGA}
\label{FPGA}
\end{center}
\end{table}

\textbf{ASIC Implementation.}
To further evaluate the performance, we implement Kaleidoscope using a 28nm ASIC library.
The implement results on ASIC are listed in the table \ref{ASIC}, where the proposed co-processor reaches 870/833/800 Mhz with 1.12/1.2/1.25ns of clock cycle.
The overall area and power-consumption of Kaleidoscope are comparable to Taurus, one of the related work on ASIC.

\begin{table}[htbp]
\begin{center}
\begin{tabular}{|c|c|c|c|}
\hline
 & \textbf{K-Base} & \textbf{K-4FPE} & \textbf{K-8FPE}\\
\hline
Freq. (MHz) & 870 & 833 & 800\\
\hline
Area (mm$^{2}$)& 3.95 & 6.42 & 10.15\\
\hline
Power (W) & 0.241 & 0.423 & 0.642\\
\hline
\end{tabular}
\caption{Kaleidoscope implementation results on 28nm ASIC}
\label{ASIC}
\end{center}
\end{table}

\subsection{Test-bed}
We establish our 100 Gbps test-bed for Kaleidoscope to evaluate the performance, shown in the figure \ref{testbed}.
A TOYO Corporation SYNESIS traffic tester is employed in out test-bed, which can both generate and replay (from pcap files) 100 Gbps of traffic.
We connect the tester and Xilinx AU-250 FPGA platform, which is mounted on the host mother board via PCIe interface.
Ubuntu 20.04 and Vivado 2020.1 are running on the host for real-time hardware debugging.
To validate the accuracy of traffic analysis, we replay two traffic analysis data-sets at the speed of the 100 Gbps via SYNESIS tester.
Inference results will be added to the optional segment in the IP header before leaving the \texttt{port 1} of the Corundum NIC.
Tester will save the returned traffic to get accuracy.

\section{Evaluation}

\subsection{Datasets and Traffic Analysis Tasks}
We conduct experiments on two traffic analysis datasets: ISCX-VPN-2016 (ISCX) \cite{ISCX} and USTC-TFC (USTC) \cite{USTC}, with three different traffic analysis tasks: encrypted traffic analysis, service-level traffic classification and malware traffic detection.
Both datasets provide pcap files of the raw traffic.
We utilize the ISCX dataset for encrypted traffic analysis tasks, which contains of 6 categories of encrypted traffic from 17 different applications.
While the USTC dataset contains traffic from 19 real-world applications with 9 different services.
Traffic from 6 common malware applications is also collected by USTC dataset.
So the service-level traffic classification and malware traffic detection is performed on the USTC dataset.

\subsection{Flexibility Evaluation and On-board NNs}

\subsubsection{On-board Raw-bytes-based NNs Models}
For three analysis tasks, Six raw-bytes-based NNs models are trained and tested on-board .
These six models includes MLP, CNNs and RNNs with various sizes, covering most common NNs applied in the traffic analysis.
MLP-E, MLP-C and MLP-M are models trained for encrypted traffic analysis, traffic classification and malware traffic detection of the mouse flows.
CNN-E, CNN-C and RNN-M are complex models trained for three tasks of the elephant flows.
To save the redundant computation, we only feeds the top valid 32/64 raw-bytes inside to above NNs, with 5 bytes for \texttt{src\_port, dst\_port, protocol} and 27/59 bytes for payload.
Experiments in Appendix A.3 discuss the impact of shorter input raw-bytes.
The information of on-board tested NNs is illustrated in the table \ref{on-board}.

\begin{table}[htbp]
\begin{center}
\begin{tabular}{|c|c|c|c|}
\hline
Object & \multicolumn{3}{c|}{NNs for the mouse flows} \\
\hline
Model & \textbf{MLP-E} & \textbf{MLP-C} & \textbf{MLP-M}\\
\hline
\textbf{Accuracy} & \textbf{92.9\%} & \textbf{97.8\%} & \textbf{98.6\%}\\
\hline
\textbf{Latency} & \textbf{352 ns} & \textbf{280 ns} & \textbf{256 ns}\\
\hline
Layer & 3 & 3 & 3\\
\hline
Size & 6.4KB & 4.4KB & 2.1KB\\
\hline
Object & \multicolumn{3}{c|}{NNs for the elephant flows} \\
\hline
Model & \textbf{CNN-E} & \textbf{CNN-C} & \textbf{RNN-M}\\
\hline
\textbf{Accuracy} & \textbf{96.1\%} & \textbf{99.2\%} & \textbf{99.3\%}\\
\hline
\textbf{Latency} & \textbf{10.2 us} & \textbf{5.7 us} & \textbf{2.7 us}\\
\hline
Layer & 5 & 5 & 4\\
\hline
Size & 280.5KB & 16.6KB & 31.4KB\\
\hline
Object & \multicolumn{3}{c|}{NNs in the related work} \\
\hline
Work & Taurus\cite{Taurus} & N3IC\cite{N3IC} & BoS \cite{bos}\\
\hline
\textbf{Accuracy} & \textbf{76.8\%} & \textbf{92.2\%} & \textbf{91.6\%}\\
\hline
NNs type & MLP & B-MLP & B-RNN\\
\hline
Size & 0.9KB & 2.5KB & 9.6KB\\
\hline
\end{tabular}
\caption{On-board tested NNs and NNs in the related work}
\label{on-board}
\end{center}
\end{table}

\subsubsection{Evaluations on Flexibility}

\begin{figure*}[htbp]
\centerline{\includegraphics[width=0.9\linewidth]{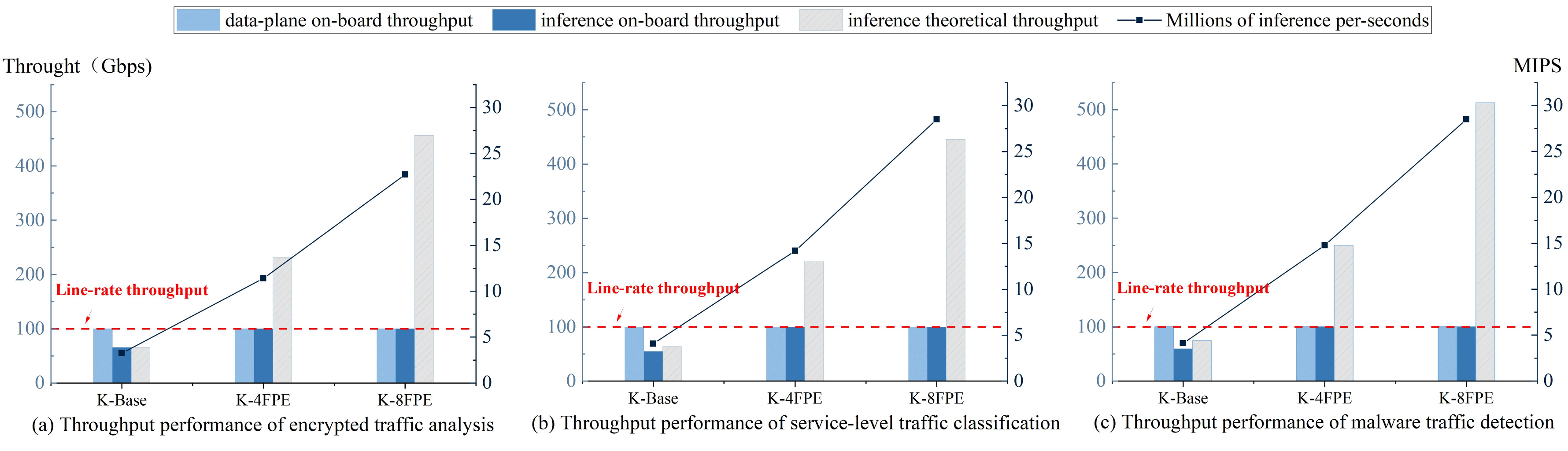}}
\caption{Throughput performance of Kaleidoscope on FPGA.}
\label{throughput}
\end{figure*}

Inferring from the table \ref{on-board}, Our Kaleidoscope architecture exhibits better flexibility from three aspects.

\textbf{Supporting NNs with different types.}
Kaleidoscope is able to support three types of NNs: MLP, CNNs, RNNs.
On contrast, existing NN-driven IDP work only support specific type of NNs.
\textbf{Moreover, it is the first time that high-accuracy CNNs model is deployed on the IDP, which is attributable to the flexibility of the RTC accelerators inside the Kaleidoscope.}

\textbf{Supporting NNs with larger sizes.}
Limited by the inline accelerators, the related work can only support very tiny NNs.
Taurus supports NNs within 1 KB (MLP in the table \ref{on-board}), while N3IC only support binary MLP (B-MLP) where all data is represented by 1-bit.
However, with memory designed in the RTC accelerator, \textbf{Kaleidoscope is able to operate the NNs that are 311.8$\times$ times larger than the related work.}
Additionally, on-board tested models occupy 3.2\%-80\% of regfile and wCache of the RTC accelerator, reflecting Kaleidoscope is possible to support even larger models.

\textbf{Supporting NNs with High-accuracy.}
Only NNs with tiny sizes or be extremely compressed can be deployed on the related work, which severely harm the accuracy of the analysis.
However, by supporting different types and larger sizes of NNs,\textbf{ Kaleidoscope achieves the SOTA accuracy among other NN-driven IDP}.
NNs on the co-processor are completed via the format of Fix-8, which maintains both hardware-friendliness and accuracy.

\subsubsection{Fix-8 Quantization}
\label{sec:5.2.3}
The Fix-8 quantization format \cite{fix8} in the paper allocates 1-bit for sign flag, 2-bit for integer and 5-bit for decimal part.
We choose Fix-8 quantization for two reasons: (i) Compared with int-8 quantization format that widely used on CPU and GPU \cite{quant1, quant2, quanttc}, Fix-8 is more hardware-friendly for saving the process of de-quantize on the FPGA; 
(ii) Compared with binary quantization used in BoS and N3IC, Fix-8 can maintain higher performance.
Unfortunately, there lacks mature and open-sourced Fix-8 library currently, and it is a difficult work to keep the differentiable characteristic for quantization-aware-training (QAT).
\textbf{We implement a Fix-8 quantization library from scratch which enables QAT to recover the accuracy of quantized models. Codes will be open-sourced.}
The detail of the development is in the Appendix A.2.

\subsection{End-to-end Performance}

\subsubsection{DP-unawareness Evaluation}
Kaleidoscope and the query table are integrated at the bypass of the data-plane. 
Before traffic leaving the forwarding pipeline, it will check this table for inference results.
Evaluation in the figure \ref{throughput} reflects that Corundum with Kaleidoscope implementation maintains 100 Gbps line-rate throughput.
Besides, the querying operation only consumes extra 5 clock cycles (15.5 ns) on the Corundum, the latency is negligible for it only adds 5.2\% latency to the original data-plane.
Such latency even cannot be detected by the traffic tester in the test-bed.
As to the functionality, the integration of Kaleidoscope modifies no function and occupies no resources inside the forwarding pipeline of Corundum, therefore maintain the original functionality of the data-plane.
The evaluation proves that Kaleidoscope co-processor introduces no significant overhead to the performance and functionality of the data-plane, achieving the DP-unawareness design goal.

\begin{figure}[htbp]
\centerline{\includegraphics[width=0.7\linewidth]{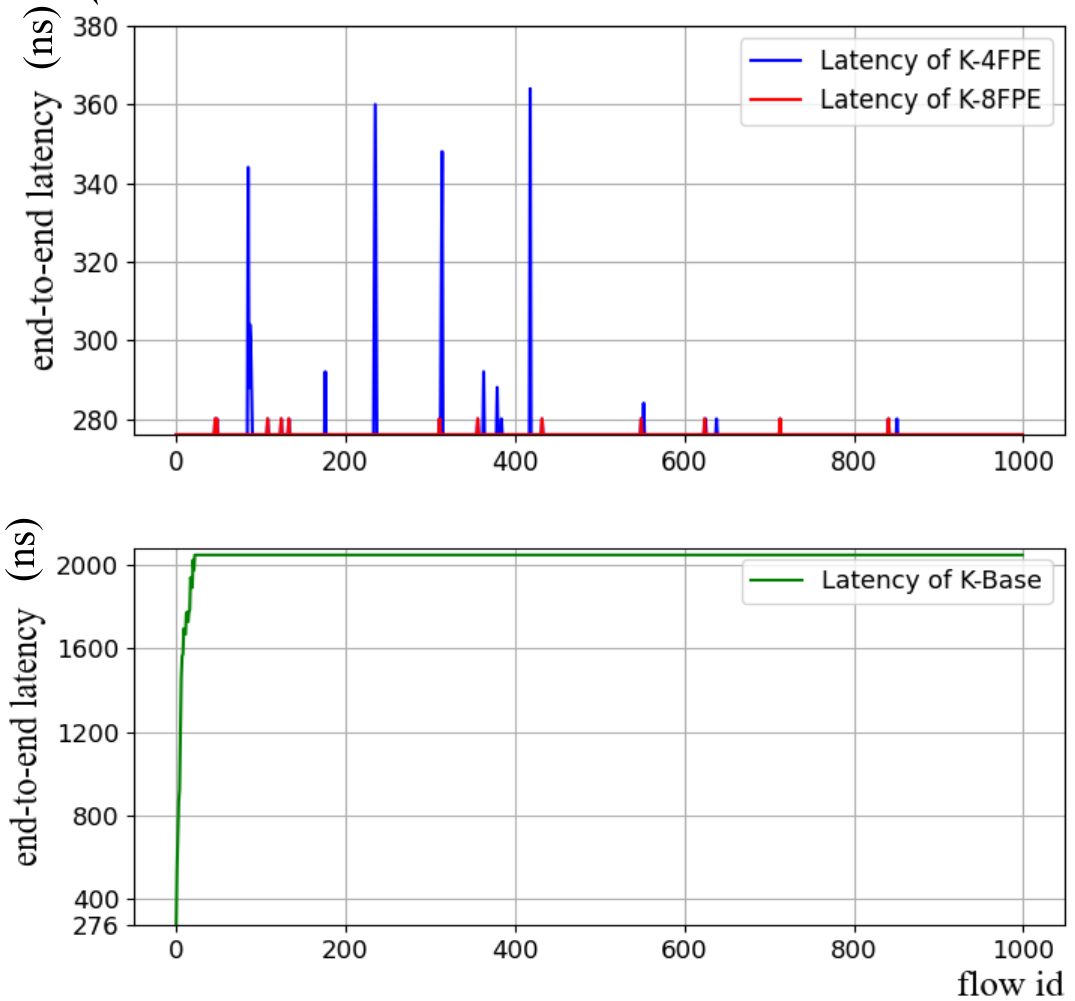}}
\caption{On-board tested latency performance.}
\label{latency}
\end{figure}

\subsubsection{Throughput Performance}
\label{sec:5.3.2}
\textbf{On-board tested throughput.}
In the figure \ref{throughput},
both K-4FPE and K-8FPE reach 100 Gbps line-rate throughput in all test-cases, achieving 3.25-28.5 millions of inference per-second (MIPS) performance, but K-Base fails to reach 100 Gbps line-rate.
The reason is in both ISCX and USTC datasets, there exist several micro-burst where the speed of concurrent flows can be over 25k flows per-second (fps).
Massive concurrent flows block the FIFO queue, leading to the loss of the packets inside the inference path.
However, with higher parallelism, K-4FPE and K-8FPE efficiently process such massive concurrent traffic.
We alas notice the packet loss inside the inference path brings no effect on the line-rate performance of the data-plane, further reflecting the DP-unawareness advantages of bypass architecture.
Three traffic analysis tasks have different traffic distribution, and we employ different NNs models for each tasks, therefore Kaleidoscope performs different throughput.

\begin{figure*}[htbp]
\centerline{\includegraphics[width=1\linewidth]{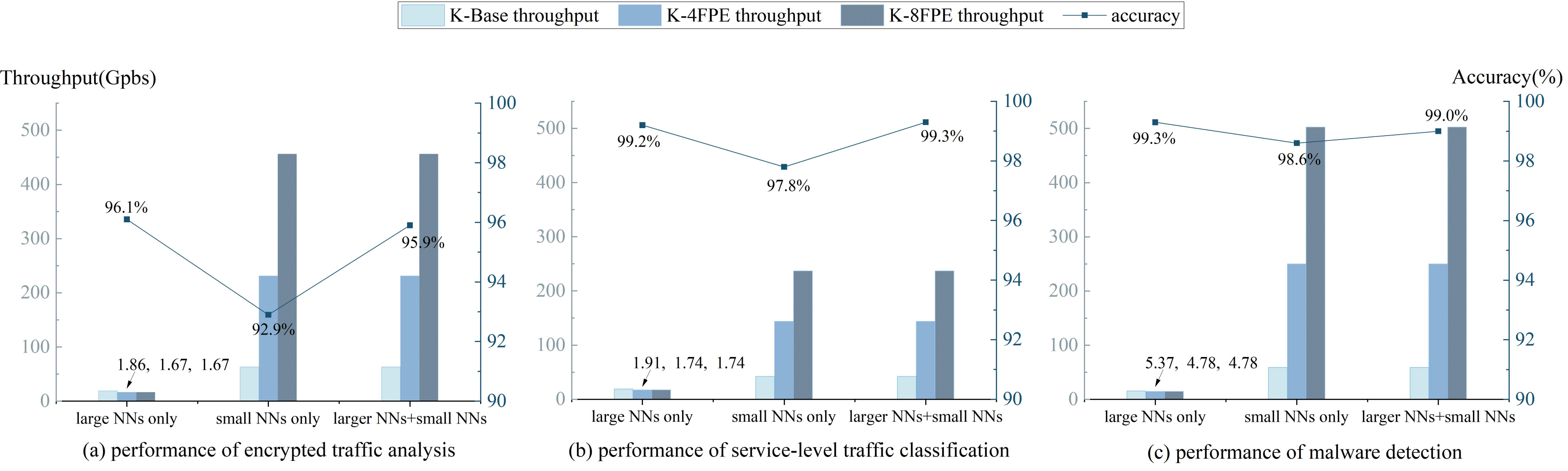}}
\caption{Performance on the elephant and mouse flows.}
\label{ablation}
\end{figure*}

\textbf{Peak performance and scalability.}
Limited by traffic tester and FPGA platforms, it only achieves 100 Gbps traffic at maximum on our test-bed.
However, there has space for K-4FPE and K-8FPE to reach higher performance.
We test the peak throughput performance of Kaleidoscope via speeding up the sending rate of flows in simulation.
Under the constraints of no packet loss inside the inference path, K-4FPE and K-8FPE acquire 87.1k fps/250.3 Gbps and 170.4k fps/502.6 Gbps peak performance in the simulation test.
Although when larger parallelism of K-4FPE and K-8FPE lowers the running frequency, the trend of performance enhancement basically follows the scaling law, proving the scalability of Kaleidoscope architecture.

\subsubsection{Latency Performance}

\textbf{Base inference latency.}
With the introduction of raw-bytes-based NNs, Kaleidoscope consumes no extra time for feature extraction, thereby the on-board tested latency is merely consisted of inference latency in RTC accelerators and queuing latency in FIFO queue.
In the table \ref{on-board}, we list the inference latency of tested NNs on FPGA.
MLP-C and MLP-E perform latency within 300 ns on FPE accelerator, meanwhile MLP-E consumes 352 ns.
This is because encrypted traffic analysis is a more challenging task, which demands larger models and longer time for completion \cite{encrypt}.
The inference latency of complex NNs for the elephant flows are also evaluated, which varies between 2.7-10.2 us on HPE accelerator.
The relative high-latency attributes to the computing complexity of larger NNs models and latency complexity of systolic array.

\textbf{End-to-end latency on board.}
We take the example of service-level traffic classification to analysis the end-to-end latency with multiple micro-burst over 25k fps.
The USTC dataset is replayed with 100 Gbps rate for around 24 seconds.
In the figure \ref{latency}, we document the on-board tested latency results of Kaleidoscope of top-1k flows in the dataset.
As analysed in section \ref{sec:5.3.2}, K-Base fails to handle the 100 Gbps traffic for full FIFO queues inside the inference path, so the incoming flows have to wait in the queue until RTC accelerators being available.
In K-Base, the inference on most flows is completed with over 2.0 us end-to-end latency, while waiting latency occupies 88.0\% of overall latency.
When it comes to the K-4FPE and K-8FPE which equipped with more RTC accelerators, latency is significantly reduced.
The micro-burst of concurrent flows still brings variance on inference latency, but in K-4FPE, 96.5\% inference finish by 280 ns without queuing, and the maximum latency is 362 ns.
While in K-8FPE, 99.5\% inference is finished by optimal 280ns.
The results proves the performance and scalability of Kaleidoscope for low-latency NNs inference on the IDP.

\subsubsection{Ablation Study on the Elephant and Mouse Flows}

In implementation, we set packet counts larger than 16 as the threshold of the mouse flows and the elephant flow.
Under this configuration, the elephant flows occupy 10.26\% of the total flows and contribute 90.15\% traffic size of the total volume in ISXC dataset, and in USTC dataset, 6.6\% of the entire flows are the elephant flows with 84.4\% traffic volume.
Small amount of elephant flows occupy most of the traffic size,
however only employing complex NNs will add inference cost and harm performance on the mouse flows.
Therefore we propose small NNs with low-latency and larger NNs with high-accuracy towards the two types of flows.

To fully evaluate the accuracy and throughput performance of our methods, we conduct ablation study on Kaleidoscope, comparing the accuracy and throughput of the small-NNs-only, large-NNs-only and small-NNs-plus-larger-NNs configurations.
The comparison is pictured in figure \ref{ablation}.
Although small NNs achieve higher throughput, they fall short on accuracy.
On contrast, when only large NNs models are deployed on the Kaleidoscope, its longer inference latency lowers throughput by 90.9\%-97.1\% on three traffic analysis tasks.
Meanwhile, when utilize small NNs for the mouse flows and large NNs for the elephant flows, it maintains the throughput performance of the small NNs, and improves the overall accuracy by 0.4\%-3.0\% for the usage of larger models.
Therefore, deploying different NNs models for the elephant flows and mouse flows contributes to achieve both high-performance and SOTA accuracy compared with other NN-driven IDP.

\subsubsection{Throughput Performance on ASIC}

\begin{figure}[htbp]
\centerline{\includegraphics[width=0.87\linewidth]{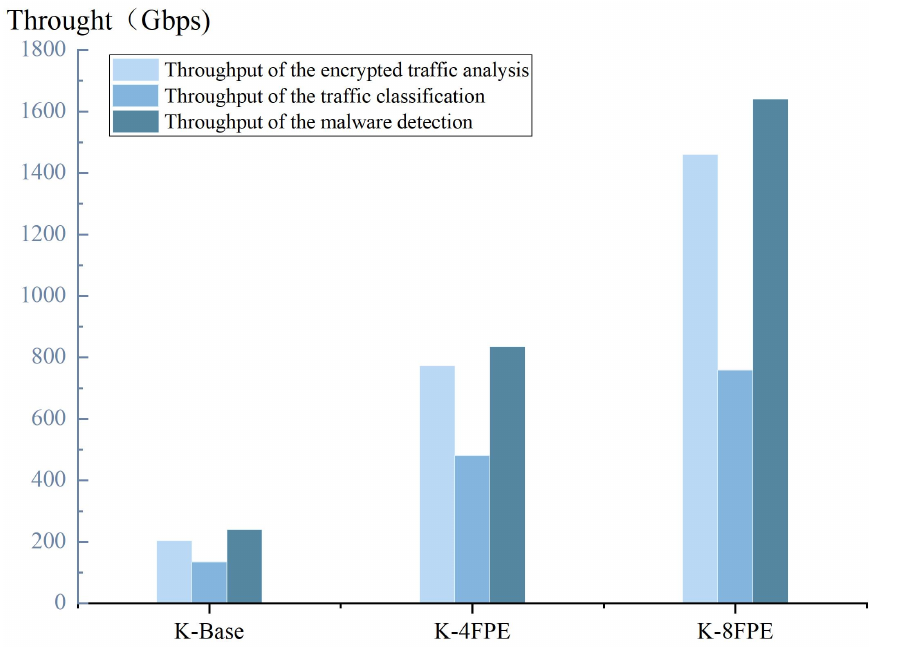}}
\caption{Throughput performance of ASIC Implementation.}
\label{ASIC_pic}
\end{figure}

The Kaleidoscope is also implemented under a 28 nm ASIC library, running at the 800-870 Mhz in the table \ref{ASIC}.
We adopt the same simulation method illustrated in section \ref{sec:5.3.2}, with evaluation performance pictured in the figure \ref{ASIC_pic}.
Kaleidoscope on ASIC reaches peak performance of 1.64 Tbps throughput.
With faster clock frequency, the inference latency of tested models are reduced to 83.2 ns and 3.1 us.

\subsection{Comparison on the Related Work}
\begin{figure}[htbp]
\centerline{\includegraphics[width=0.82\linewidth]{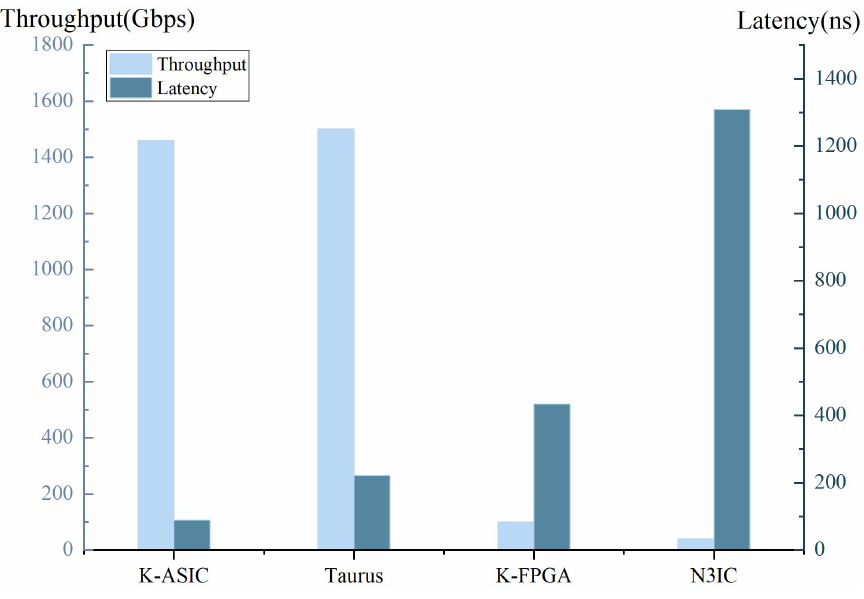}}
\caption{Performance comparison on the related work.}
\label{related}
\end{figure}

We choose Taurus and N3IC for comparison, both of which are latest hardware solutions for NN-driven IDP.
Taurus is implemented on the ASIC with 15 nm library and 1 GHz running frequency, while the N3IC is implemented on NetFPGA \cite{netfpga}.
For fair comparison, we contrast Taurus with Kaleidoscope on ASIC (K-ASIC), and contrast FPGA-based Kaleidoscope (K-FPGA) with N3IC.
Besides, we deploy the NNs utilized in the Taurus or N3IC in this part of evaluation.
The comparing results are in the figure \ref{related}.

Kaleidoscope and Taurus both reach throughput over 1.4 Tbps. Although running at the higher frequency, Taurus works with 221 ns inference latency, which is 2.5$\times$ higher than the K-ASIC.
The reason is Taurus adopts the inline architecture to integrating accelerator and the forwarding pipelines, which increases the unavoidable latency for feature extraction.
However, K-ASIC deploys raw-bytes-based NNs, saving the extra extraction time.
On the other hand, both K-FPGA and N3IC achieves the line-rate performance on FPGA.
However, K-FPGA outperforms N3IC on latency by 2.3 $\times$ lower.
To conclude, Kaleidoscope exceeds the existing hardware solutions on inference performance.

Kaleidoscope also outperforms N3IC and Taurus on flexibility.
Both work designs an inline pipelined accelerator for the data-plane, where NNs flexibility is severely limited.
In the table \ref{on-board}, the largest NNs supported by Taurus and N3IC is within 2.5 KB, while Kaleidoscope can efficiently support NNs over 200 KB.
Besides, related work can only complete inference with specific types of NNs, while in the evaluation, six NNs models including MLP, CNNs and RNNs are tested on-board.
The flexibility characteristic contributes to the higher accuracy compared with Taurus and N3IC.
As the comparison in the table \ref{on-board}, in malware detection tasks, Kaleidoscope acquires 26.5\% and 7.7\% higher accuracy.

\section{Discussion and Future Work}

\textbf{Traffic extractor.}
Although raw-bytes-based NNs are utilized in this paper, in the future, we still plan to support feature-based NNs on Kaleidoscope by implementing hardware sketch \cite{sketch1, sketch2, pcsketch} as traffic extractor at the bypass.


\textbf{Hash collision in the flow table.}
For rapid development, the hash module inside the Corundum NIC is reused in Kaleidoscope, which may introduce hash collision in the flow table.
There exists solutions for the problem \cite{hash, hashtc} such as calculating multiple hash functions simultaneously, which is feasible in the Kaleidoscope.
We leave it for the further improvement.


\textbf{Discussion on high-performance.}
It is worthwhile to analysis why our RTC accelerators can outperform or achieve comparable performance on contrast to the pipelined accelerator.
The reason is two-fold.
First, Kaleidoscope is located at the bypass and raw-bytes-based NNs are introduced, which save latency of waiting forwarding pipelines and feature extraction. 
This reflects the advantages of our architecture on low-latency.
The low-latency optimizations in FPE accelerator contribute to the performance as well.
Secondly, the proposed RTC accelerators perform comparable throughput by ensuring the non-stall data flow for computation via VLIW and memory design.
These techniques makes Kaleidoscope acquires comparable throughput with the pipeline architecture.

\section{Conclusion}
In this paper we propose Kaleidoscope, a flexible and high-performance co-processor for NN-driven IDP.
Kaleidoscope adopts a unique bypass architecture with programmable RTC accelerators, high-performance inference engine towards elephant/mouse flows and raw-bytes-based NNs.
We evaluate the design of Kaleidoscope on FPGA-based 100 Gbps test-bed and a 28 nm ASIC library. 
Evaluation results proves Kaleidoscope meets the design goal of flexibility, high-performance and DP-unawareness.

{\footnotesize \bibliographystyle{acm}
\bibliography{sample}}


\end{document}